\documentclass[twocolumn,preprintnumbers,amsmath,amssymb,superscriptaddress]{revtex4-2}
\UseRawInputEncoding

\usepackage{latexsym,amssymb,amsthm,amsmath,epsfig,braket}
\usepackage[left=2.00cm, right=2.00cm, top=2.00cm, bottom=2.00cm]{geometry}
\usepackage{xcolor}
\usepackage{svg}
\usepackage[colorlinks, citecolor={blue!80!black}, urlcolor={blue!80!black}, linkcolor={red!50!black}]{hyperref}
\usepackage{float}
\usepackage{hyperref}
\hypersetup{
	citecolor=blue,
	colorlinks=true,
	linkcolor=blue,
	filecolor=blue,      
	urlcolor=blue,
}

\begin{document}
\title{Loss-free enhancement of photonic spin Hall shift by electromagnetically induced transparency}
\author{Kezhou Du}
\thanks{These authors contributed equally to this work.}
\author{Aizaz Khan}
\thanks{These authors contributed equally to this work.}
\affiliation{School of Physical Science and Technology, Soochow University, Suzhou 215006, China \& Jiangsu Key Laboratory of Frontier Material Physics and Devices, Soochow University, Suzhou 215006, China.}
\author{Lei Gao}
\affiliation{School of Optical and Electronic Information, Suzhou City University \& Suzhou Key Laboratory of Biophotonics \& Jiangsu Key Laboratory of Biophotonics, Suzhou 215104, China}
\author{Muzamil Shah}
\email{muzamill@qau.edu.pk}
\affiliation{Department of Physics, Quaid-e-Azam University, Islamabad, Pakistan}
\author{Xinxing Zhou}
\email{xinxingzhou@hannu.edu.cn}
\affiliation{Key Laboratory of Low-Dimensional Quantum Structures and Quantum Control of Ministry of Education, School of Physics and Electronics, Hunan Normal University, Changsha 410081, China}
\author{Dongliang Gao}
\email{dlgao@suda.edu.cn}
\affiliation{School of Physical Science and Technology, Soochow University, Suzhou 215006, China \& Jiangsu Key Laboratory of Frontier Material Physics and Devices, Soochow University, Suzhou 215006, China.}

\begin{abstract}
The photonic spin Hall effect (PSHE), a result of spin-orbit interaction, has attracted significant interest because of its fundamental importance and potential applications. Optical losses are ubiquitous, which inherently suppress the photonic spin Hall shift (PSHS). In this work, we consider an atomic medium that exhibits both absorption and transparency to investigate and mitigate the effects of loss on PSHS. We demonstrate that laser-induced coherence in an atomic medium, leading to electromagnetically induced transparency (EIT) at resonance, counteracts the detrimental effects of losses on the PSHS. Upon EIT in a coherent medium enclosed within dielectric slabs, the reflectivity of the incident polarized state is reduced near Brewster's angle to enhance PSHS. Moreover, the tunable refractive index of the atomic medium enables the manipulation of PSHS without structural modifications with a tiny loss. Our proposed loss-free approach to PSHS may enable advanced optical sensing and other spin-based applications.
\end{abstract}
\maketitle

\section{Introduction} \label{sec:outline}

Light exhibits wave-particle duality and, like classical particles, carries angular momentum comprising intrinsic spin angular momentum (SAM) stemming from the polarization and orbital angular momentum (OAM) originating from spatial degrees of freedom. The SAM is defined by the polarization helicity, while the OAM is related to the beam's azimuthal phase variation, featured with its topological charge \cite{bliokh2010angular}. Conservation of angular momentum leads to the coupling of SAM and OAM, i.e., the spin-orbit interaction (SOI) of light and provides a platform to manipulate the light-matter interaction \cite{pan2016strong,bliokh2015spin,tsesses2019spin,cui2021probing,rodriguez2010optical,hosten2008observation}. Interestingly, the SOI can lead to a wavelength-scale shift of reflected light known as the photonic spin Hall shift (PSHS), which has attracted enormous scientific interest due to its fundamental significance and practical applications \cite{mi2017precise,kort2017topological,zhou2018broadband}. However, the resulting PSHS is quite diminutive owing to the limited extent of SOI, which necessitates specialized methods to enhance the tiny shift, such as quantum weak measurement \cite{hosten2008observation,zhou2012experimental} and multiple reflections \cite{bliokh2008geometrodynamics,nori2008dynamics}. Consequently, a great deal of interest has been given to enhancing the spin splitting in epsilon-near-zero media \cite{chen2022wide}, two-dimensional materials \cite{bai2017tunable,kort2018photonic,wu2020actively}, plasmonic materials \cite{yu2021spin, wan2020controlling} and nanoparticles \cite{sun2022wavelength,olmos2019enhanced,gao2018enhanced,olmos2023optical}.

For an incident \textit{p}-polarized light at an air-glass interface, the PSHS is proportional to the ratio $r_{s} /r_{p} $ \cite{luo2011enhanced}. Therefore, PSHS can be effectively enhanced by suppressing the Fresnel's coefficient of the incident polarization state \cite{kim2023spin}. For example, at an air-glass interface near Brewster's angle, PSHS for \textit{p}-polarized incident light can be enhanced \cite{ling2021revisiting}. However, in practical scenarios, optical systems are subjected to material losses, which is detrimental to the enhancement of PSHS. In order to mitigate the effect of a lossy medium, optical systems are loaded with some gain counterparts such as non-Hermitian systems with balanced gain and loss \cite{miri2019exceptional}. Such systems exhibit exceptional points regardless of incident polarization and giant PSHS can be obtained as the system transitions through different phases \cite{zhou2019controlling}. Attaining exceptional points in a balanced gain-loss system requires a high contrast of gain and loss, which is hard to achieve practically. Considering this, other strategies must be implemented to realize the enhancement of PSHS.

Quantum interference, which arises as consequences of the laser-induced coherence between various atomic states, has generated numerous phenomena of practical interest \cite{fleischhauer2005electromagnetically}. 
Coherent preparation of quantum states in atoms or molecules can introduce quantum interference in the amplitudes of optical transitions. As a result, the optical properties of a medium can be significantly altered with quantum interference \cite{budker1999nonlinear,bajcsy2003stationary,hu2004enhanced}. Among these phenomena, electromagnetically induced transparency (EIT) has established the gas-phase systems as a leading platform for realizing complete transparency in media. By leveraging quantum interference between transition pathways, systems achieve no dissipation (absorption) of the field at the resonant frequency of optical transitions \cite{harris1990nonlinear,rohlsberger2012electromagnetically,mucke2010electromagnetically}. In the vicinity of the resonance frequency of the EIT window, normal dispersion controlled by the intensity of the driving field emerges, though the refractive index passes through the vacuum value \cite{yadav2019switching}. Recently, optical systems composed of a cavity containing atomic gases under EIT conditions have been extensively studied to control the related optical phenomena known as Goos-H\"anchen \cite{wang2008control,ziauddin2010coherent} and Imbert-Fedorov shifts \cite{fedorov2013theory, imbert1972calculation, asiri2016controlling}. In this work, we demonstrate that the cavity filled with atomic gases exhibiting EIT at resonance frequency, enclosed by dielectric walls, can reduce the effects of losses (absorption) on the PSHS. The complete elimination of absorption leads to a dramatic enhancement of the magnitude of the PSHS by several orders without introducing compensatory gain. The restoration of the absorption peak at resonance with anomalous dispersion sufficiently suppresses the magnitude of PSHS. In addition to loss-free PSHS, the tunable refractive index of the atomic medium can dynamically tune PSHS with low-absorption by detuning from resonance. These findings establish a reliable pathway to applications in spin-based refractive index sensing.

\section{Theory Analysis}
Our proposed model is shown in Fig. \ref{FIG:1}, which can be described as a trilayered structure forming a cavity to host a four-level \textit{N}-type medium in the middle of the two non-magnetic dielectric layers. For simplicity, the structure is kept in vacuum with a dielectric function $\varepsilon_{0} =1$. The non-magnetic dielectric layers share a common permittivity $\varepsilon_{s}$ and thickness $d_{s}$, whereas a relatively thick coherent medium is characterized by the thickness $d_{a}$, permittivity $\varepsilon_{a}$ and permeability $\mu _{a}$. The permittivity of the slabs $\varepsilon_{s}$ is chosen 3, whereas the permittivity of the medium inside the cavity is given by the relation $\varepsilon_{a} =1+\chi$, with $\chi$ being the susceptibility of the medium. The configuration forms a slab-cavity-slab (SCS) trilayered structure which is illuminated by a probe field of wavelength $\lambda$ at an incident angle $\theta_{i}$ relative to the positive \textit{z}-axis (normal to the interface). The reflectance and transmittance can be derived using the transfer matrix as \cite{asiri2016controlling}:
\vspace{0.5pt}
\begin{figure}[H]
	\centering
	\includegraphics[width=\columnwidth]{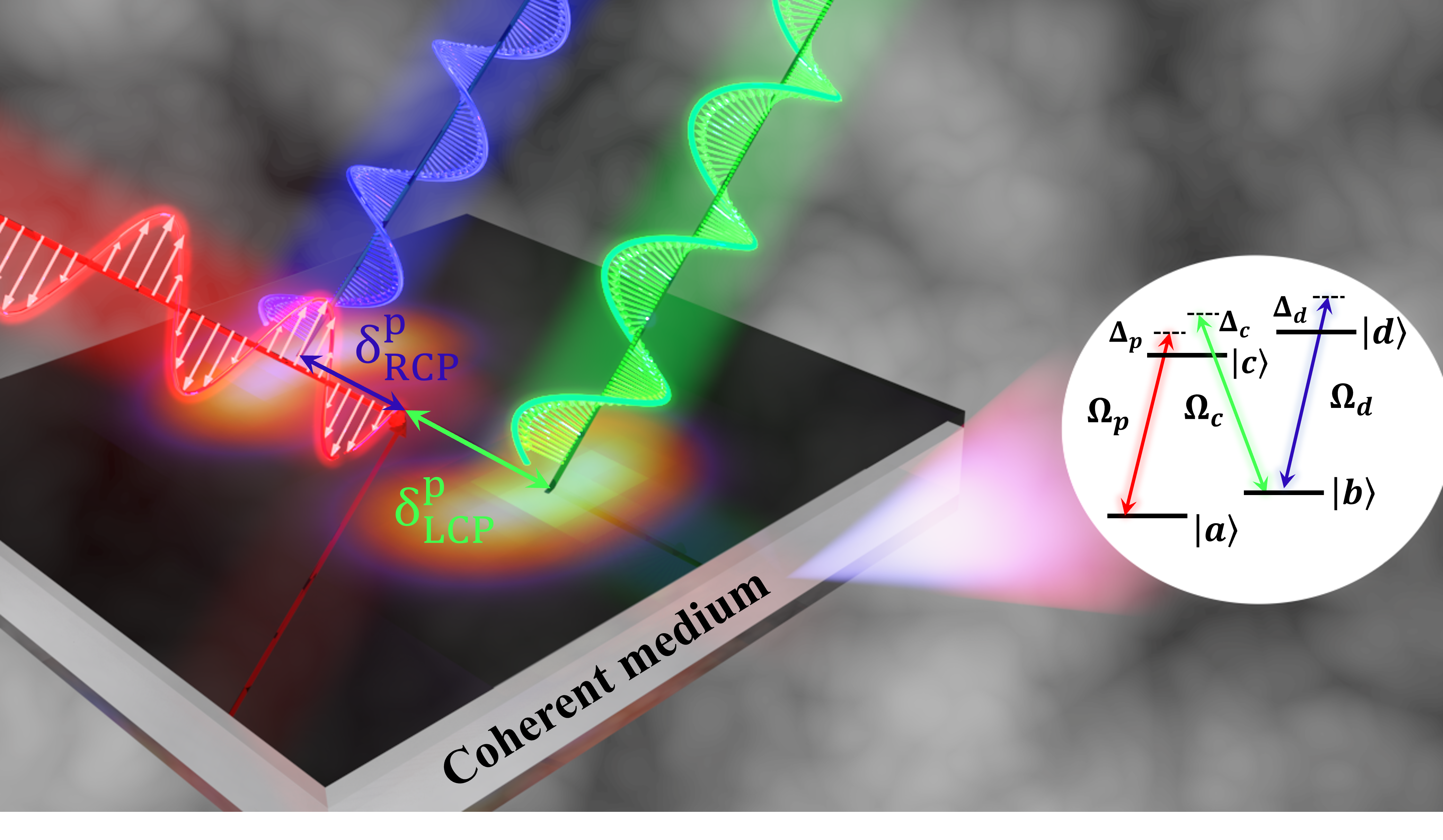}
	\caption{Schematics of the proposed model, containing an \textit{N}-type coherent atomic medium forming a cavity enclosed by dielectric slabs. A linearly polarized incident light suffers spin-splitting into its constituent RCP and LCP components upon reflection, each shifted by a transverse displacement of equal and opposite sign. The beam centroids of the reflected RCP and LCP components are shown on the interface of air and dielectric. The inset shows various transitions between atomic states in the coherent atomic medium.}
	\label{FIG:1}
\end{figure}

\begin{widetext}
\begin{equation}\label{eq1}
	M_{j}^{p/s} (\theta _{i} ,d_{j} )=\left(\begin{array}{cc} {\cos (\eta _{j} k\cos (\theta _{j} )d_{j} )} & {\frac{i\alpha_{j}^{p/s} \sin (\eta _{j} k\cos (\theta _{j} )d_{j} )}{\eta _{j} \cos (\theta _{j} )} } \\ {\frac{i\eta _{j} \cos (\theta _{j} )\sin (\eta _{j} k\cos (\theta _{j} )d_{j} )}{\alpha _{j}^{p/s} } } & {\cos (\eta _{j} k\cos (\theta _{j} )d_{j} )} \end{array}\right),\end{equation}
\end{widetext}
where $k$ is the free-space wavevector, $\theta _{j}$ is the refraction angle for each interface and can be related to the incident angle $\theta _{i}$ according to Snell's law, $\eta_{i}\sin(\theta_{i})=\eta_{j}\sin(\theta _{j})$. Here, $\eta_{j} $ and $d_{j}$ are the refractive index and the thickness of the \textit{j${}^{th}$} layer, and the parameter $\alpha _{j}^{p/s}$ depends upon the incident polarization, that is, $\alpha_{j}^{p}=\varepsilon_{i} $ and $\alpha_{j}^{s} =\mu _{i} $. The proposed model can be considered as a two-port SCS structure where the complete transfer matrix is formulated as a matrix of order 2, as follows \cite{wang2008control}:

\begin{equation}\label{eq2}
	Q^{p/s} (\theta _{i} ,d_{j} )=M_{s}^{p/s} (\theta _{i} ,d_{s} )M_{a}^{p/s} (\theta _{i} ,d_{a} )M_{s}^{p/s} (\theta _{i} ,d_{s} ),  
\end{equation}
\\
where each $M_{j}^{p/s} (\theta _{i} ,d_{j} )$ depends on the parameters of the corresponding layer and the superscript \textit{p/s} represents the incident wave polarization. The reflectance $R^{p/s} (\theta _{i},d_{j} )$ and transmittance $T^{p/s} (\theta _{i} ,d_{j} )$ for the whole cavity can be obtained from the Fresnel's coefficients as defined by Eq. \eqref{eq9} and Eq. \eqref{eq10} in the Appendix, respectively. The absorptance can be explicitly calculated by energy conservation, that is, $A^{p/s} (\theta _{i} ,d_{j} )=1-R^{p/s} (\theta _{i} ,d_{j} )-T^{p/s} (\theta _{i} ,d_{j} )$.

We consider a Gaussian beam illuminates on a planar interface whose angular spectrum is given by;

\begin{equation}\label{eq3a}
\tilde{E}_{i} =\frac{w}{2\pi}e^{-\frac{w^{2} (k_{ix}^{2} -k_{iy}^{2} )}{4}}.
\end{equation}
\\ 
Here, $w$ denotes the beam-waist, while $k_{ix} $ and $k_{iy} $ represent the \textit{x} and \textit{y}-components of the incident wavevector \cite{ling2021revisiting,zhou2019controlling}. Using Fresnel's coefficients obtained above, the reflected fields can be calculated by enforcing the boundary conditions as \cite{bliokh2007polarization}:
\\
\begin{equation}\label{eq3}
	\left( 
	\begin{array}{c}
		{\tilde{E}}^{p}_r \\ 
		{\tilde{E}}^{s}_r
	\end{array}
	\right)=\left(
	\begin{array}{cc}
		r^{p} & \frac{k_{ry}cot (\theta_i)(r^{p}+r^{s})}{k} \\ 
		-\frac{k_{ry}cot (\theta_i)(r^{p}+r^{s})}{k} & r^{s} 
	\end{array}
	\right)\left( \begin{array}{c}
		{\tilde{E}}^{p}_i \\ 
		{\tilde{E}}^{s}_i \end{array}
	\right),
\end{equation}
\\
where $r^{p/s}$ represents the Fresnel's coefficients. Additionally, the \textit{z}-component of the electric field is small enough to neglect. As to the reflected fields, the angular spectrum $\tilde{E}_{r}^{+}$ and $\tilde{E}_{r}^{-}$ can be calculated for individual LCP and RCP components. The beam centroids for LCP and RCP are obtained from their respective angular spectrums using the following relation \cite{zhou2019controlling}:
\\
\begin{equation}\label{eq3b}
\left\langle \delta _{LCP/RCP} \right\rangle =\frac{{\left\langle \tilde{E}_{r\pm }  \right|} i\partial\boldsymbol k_{ry} {\left| \tilde{E}_{r\pm }  \right\rangle} }{{\left\langle \tilde{E}_{r\pm }  \right|} \left. \tilde{E}_{r\pm } \right\rangle },
\end{equation}
\\
where $\partial\boldsymbol k_{ry} =(\partial /\partial k_{ry} )\boldsymbol{\mathrm{e}}_{ry} $. 
Considering the incident wave of \textit{p}-polarization, the PSHS $\delta_{LCP/RCP}^p$ of the reflected light can be calculated as follows:

\begin{equation}\label{eq4}
	\delta _{LCP/RCP}^{p}=\mp \frac{k_{0} w^{2} |r^{p} |^{2}(1+|r^{s} |/|r^{p} |\phi)\cot (\theta _{i} )}{k_{0}^{2} w^{2} |r^{p} |^{2} +|\partial r^{p} /\partial \theta _{i} |^{2} +\xi\cot ^{2} (\theta _{i} )},  
\end{equation}
\\
where $\xi =|r^{p} |^{2} +|r^{s} |^{2} +2|r^{p} ||r^{s} |\phi$, $\phi=\cos (\varphi _{s} -\varphi _{p} )$ with $\varphi _{p/s}$ being the phase of the reflection coefficients, i.e., $r^{p/s} =|r^{p/s} |e^{i\varphi _{p/s} }$. The above equation can be used to determine the PSHS of incident \textit{p}-polarized wave.

\section{Optical Susceptibility of the four-level intra-cavity medium}
In our proposed model, the coherent medium which forms the cavity can be taken any generic levels system that supports EIT. However, to obtain both EIT and electromagnetically induced absorption (EIA) like regimes, we proceed with a four levels \textit{N}-type atomic system \cite{yadav2019switching}, as shown in the inset of Fig.\ref{FIG:1}. A weak probe field defined by the Rabi frequency $\Omega _{p} =\sigma _{ac} E_{p} /2\hbar $ 
coupling the transition ${\left| a \right\rangle} \to {\left| c \right\rangle} $ is detuned from the atomic transition frequency $\omega _{ca} $ by a factor $\Delta _{p} =\omega _{p} -\omega _{ca} $. Additionally, the medium is driven by strong control fields defined by Rabi frequency $\Omega _{c,d} =\sigma _{ij} E_{i} /2\hbar $, coupling the transitions ${\left| b \right\rangle} \to {\left| c \right\rangle} $ and ${\left| b \right\rangle} \to {\left| d \right\rangle} $, respectively. The corresponding detuning is defined as $\Delta _{i} =\omega _{i} -\omega _{ij} $ and $\sigma _{ij}$ is the corresponding electric dipole moment of transitions. Within the framework of the electric dipole and rotating-wave approximations, the system's Hamiltonian of interaction picture takes the form as \cite{wan2020controlling}:

\begin{align}\label{eq5}
	H_{i}&=-\hbar \big((\Delta _{p} -\Delta _{c} ){\left| b \right\rangle} {\left\langle b \right|} +\Delta _{p} {\left| c \right\rangle} {\left\langle c \right|} +(\Delta _{p} -\Delta _{c} +\Delta _{d} )\notag \\
	&{\left| d \right\rangle} {\left\langle d \right|}+(\Omega _{p} {\left| c \right\rangle} {\left\langle a \right|} +\Omega _{c} {\left| c \right\rangle} {\left\langle b \right|} +\Omega _{d} {\left| d \right\rangle} {\left\langle b \right|} +H.c)\big), 
\end{align}
\\
where $H.c$ is the Hermitian conjugate. The population and coherence dynamics are governed by the time evolution of the density matrix, expressed through the Liouville equation written in the terms of the commutation of the Hamiltonian and density operator, i.e., $\dot{\rho} =-(i/\hbar )[H_{i} ,\rho ]+L_{p}$ \cite{wan2020controlling}. Here, $\rho$ represents the density matrix, whose diagonal elements describe the populations while off-diagonal elements describe the coherence. The second term $L_{p}$ governs the relaxation process. By inserting $H_{i} $ in the Liouville equation and considering various decay processes, one can obtain the rate equation for $\dot{\rho}$. The matrix elements $\dot{\rho _{ij}}={\left\langle i \right|} \dot{\rho} {\left| j \right\rangle} $, as given in the Appendix, are constraints to the population conservation i.e., $\rho _{aa} +\rho _{bb} +\rho _{cc} +\rho _{dd} =1$, and $\rho _{ij} =\rho _{ji}^{*} $ for $i\ne j$ situation, where $\Gamma _{ji} $ is the decay rate for the corresponding transitions and $\gamma _{ba}$ is the decoherence between the ground states ${\left| b \right\rangle} $ and ${\left| a \right\rangle} $.

\begin{figure*}
	\centering
	\includegraphics[width=\textwidth]{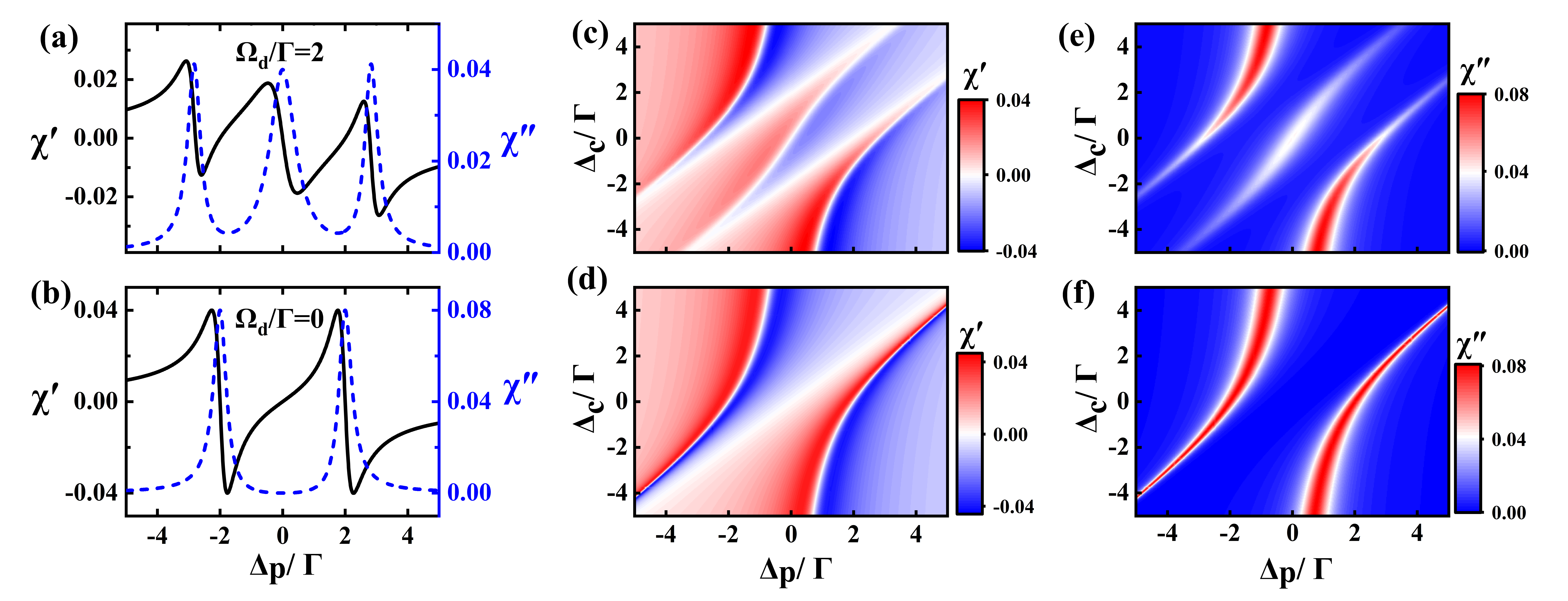}
	\caption{ Complex susceptibility as a function of detuning $\Delta _{p}$ from resonance (a) in the presence of $\Omega _{d}$ and (b) absence of $\Omega _{d}$. The susceptibility of the atomic medium exhibits distinct optical responses depending on the presence of the control field $\Omega _{d}$. The optical response shows high absorption at resonance backed by steep anomalous dispersion in the presence of $\Omega _{d}$. In the absence of $\Omega _{d}$, normal dispersion appears at resonance which is associated with the emergence of a transparency window. (c, d) The real part and  (e, f) imaginary part of the susceptibility in the presence (first row) and absence (second row) of the control field $\Omega _{d}$, respectively, as a function of detunings from resonance of probe $\Delta _{p}$ and control $\Delta _{c}$ field. The parameters of this structure are fixed at $\beta=0.04\Gamma$, $\gamma _{ba}=0$, $\Gamma_c=\Gamma_d=\Gamma$ and $\Delta_d=0\Gamma$. }
	\label{FIG:2}
\end{figure*}

Considering the steady-state solution of the rate equations under weak-probe conditions, the polarization of the medium can be expressed in terms of the coherence  $\rho _{ca} $ as $P=N\sigma _{ca} \rho _{ca} $, where $N$ denotes the atomic number density. Furthermore, this polarization determines the susceptibility $\chi$ of the medium via the equation $P=\varepsilon _{0} \chi E_{p} $. By equating both definitions and using $E_{p} =2\hbar \Omega _{p} /\sigma _{ca} $, one can derive the susceptibility $\chi $. After simplification, the resulting susceptibility of the intra-cavity medium has the following form \cite{wan2020controlling}:

\begin{equation}\label{eq6}
	\chi =-\beta \frac{\xi _{ba} \xi _{da} -\Omega _{d}^{2} }{\xi _{ca} (\xi _{ba} \xi _{da} -\Omega _{d}^{2} )-\xi _{da} \Omega _{c}^{2} },    
\end{equation}
\\
where $\xi _{ba} =(\Delta _{p} -\Delta _{c} )+i\gamma _{ba} $, $\xi _{ca} =\Delta _{p} +i\Gamma _{c}/2$, $\xi _{da} =(\Delta _{p} -\Delta _{c} +\Delta _{d} )+i\Gamma _{d} /2$ and $\beta =N\sigma _{ca}^{2} /2\hbar \varepsilon _{0} $. $\Gamma _{c}=\Gamma _{ca} +\Gamma _{cb}$, $\Gamma _{d}=\Gamma _{da} +\Gamma _{db}$ represent the total decay rates from levels ${\left| c \right\rangle}$ and ${\left| d \right\rangle}$. The permittivity of the medium $\varepsilon _{a} =1+\chi $ is thus directly derived from the susceptibility. By adjusting the intensity of the driving fields and the corresponding detunings, the permittivity of the intra-cavity medium can be precisely controlled. The possibility of tuning the medium's permittivity provides a direct mechanism for controlling the PSHS, which will be discussed in later sections.

\section{Results and Discussions}
In this section, we analyze the PSHS of the reflected beam while considering \textit{p}-polarized incident light. For \textit{s}-polarized incident light, the spin-dependent shift is negligible and excluded from our discussion. Before delving into the spin-dependent splitting of the reflected light, we focus on the optical response of the medium inside the neutral slabs (non-lossy). The interaction of the medium with the probe field can be analyzed through its polarization or susceptibility as defined by Eq. \eqref{eq6}. The optical susceptibility decomposes into real and imaginary components, governing the dispersion and absorption of the weak probe field, respectively. Under weak probe fields, the linear response of the susceptibility suffices to explain the intracavity medium's permittivity.

To study PSHS at resonance conditions in the presence of absorption or otherwise complete transparency, we first discussed the susceptibility of the medium which describes dispersion and absorption by real and imaginary parts, respectively. In Fig. \ref{FIG:2}, we compared the medium's optical response in the presence and absence of the strong control field $\Omega _{d}$ which drives the transition ${\left| b \right\rangle} \leftrightarrow {\left| d \right\rangle}$, and we neglect the decay $\gamma _{ba}$ between the ground state. The cavity with \textit{N}-type systems can be obtained with the $D_2$ transition at wavelength $\lambda=589.1$ nm of the probe field. It is worth noting that the imaginary part of the susceptibility exhibits a peak at resonance similar to EIA in the presence of $\Omega _{d}$, where light experiences steep anomalous dispersion as shown in Fig. \ref{FIG:2}(a). The peak absorption reduces when the detunings of the weak probe match the detuning of the control field. From Eq. \eqref{eq6}, one can easily guess this from the first term in the numerator, i.e., $\xi _{ba}$ approaches zero. This effectively reduces the susceptibility as shown in Figs. \ref{FIG:2}(c,e). Next, we turn off the strong control field $\Omega _{d}$ and assume that the control field $\Omega _{c}$ is at resonance with the transition frequency $\omega_{bc}$. Under these assumptions, the real ($\chi'$) and imaginary ($\chi''$) parts of the susceptibility reduces to the following form:

\begin{equation}\label{eq7}
	\chi '=-\frac{4\beta \Delta _{p} (\Omega _{c}^{2} -\Delta _{p}^{2} )}{4\Delta _{p}^{4} +4\Omega _{c}^{4} +\Delta _{p}^{2} (\Gamma _{c}^{2} -8\Omega _{c}^{2} )} ,
\end{equation}
\\
\begin{equation}\label{eq8}
	\chi ''=\frac{2\beta \Gamma _{c} \Delta _{p}^{2} }{4\Delta _{p}^{4} +4\Omega _{c}^{4} +\Delta _{p}^{2} [\Gamma _{c}^{2} -8\Omega _{c}^{2} ]} .    
\end{equation}
\\
It is straightforward from Eq. \eqref{eq8} that the absorption $\chi''$ vanishes at resonance $\Delta _{p} =0$, leading to a transparency window, as shown in Fig. \ref{FIG:2}(b). In the absence of the driving field $\Omega _{d}=0$, the system becomes similar to $\Lambda $-type medium, which at $\Delta _{c} =0$, leads to EIT. The off-resonant absorption peaks can be explained by the formation of the dressed states \cite{scully1997quantum}. Remarkably, the presence of a coherent control field $E_{c} $ induces EIT in an otherwise thick and opaque medium. Moreover, the light suffers normal dispersion at resonance. Next, we remove the constraint of the resonant driving field $E_{c}$ and showed that the transparency window can be precisely tuned by enforcing the conditions $\Delta _{p}=\Delta _{c}$, as shown in Figs. \ref{FIG:2}(d,f). Turning on the control field $\Omega _{d}$ redistributes the population, and the transparency window disappears at the resonance while anomalous dispersion is obtained with high absorption. Under induced transparency, we will show that the PSHS is enhanced in contrast to the presence of absorption.

\begin{figure}
\centering
\includegraphics[width=\columnwidth]{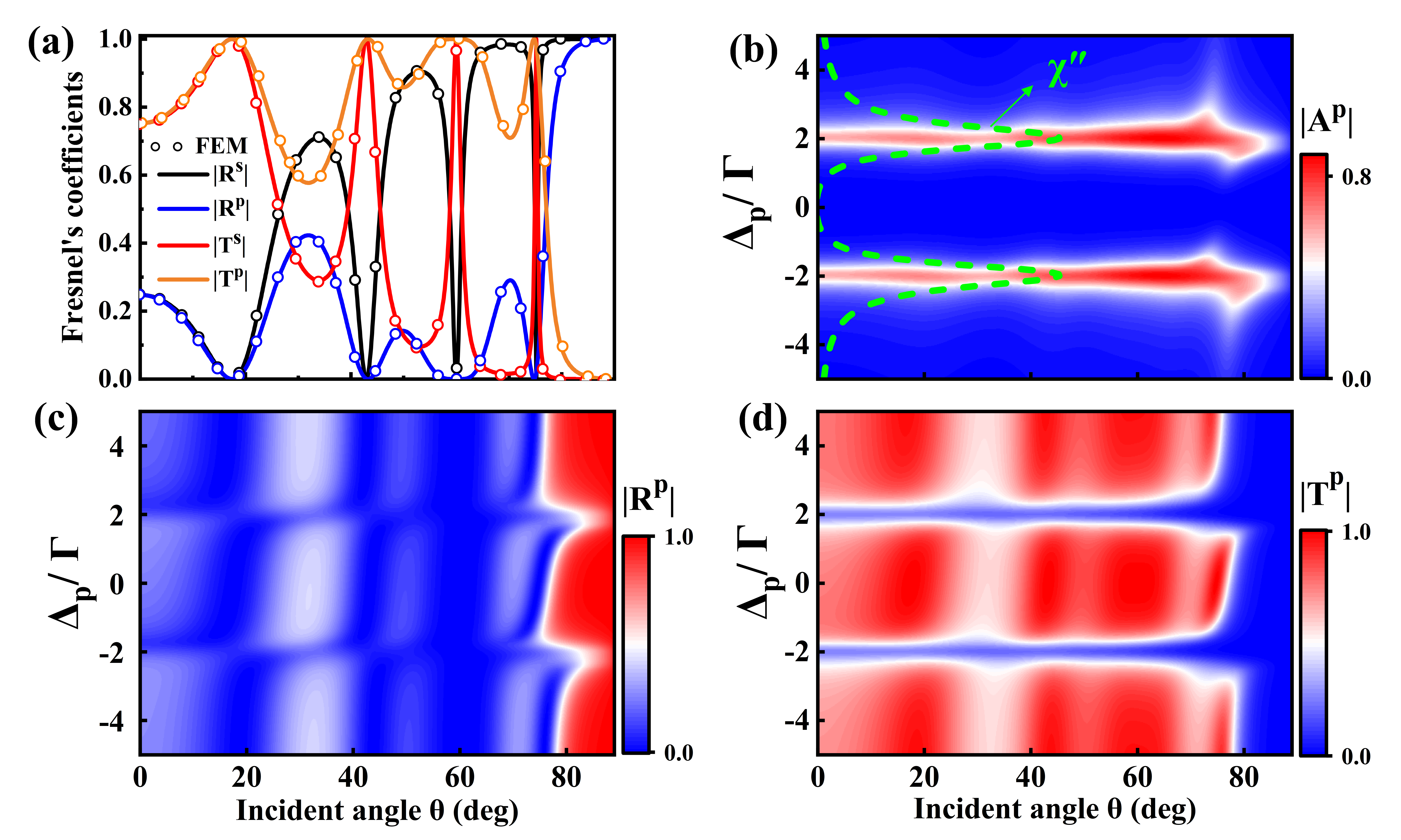}
\caption{Optical response of the whole slab-cavity-slab structure. (a) Reflectance and transmittance as functions of the incident angle. The reflectance of the cavity approaches zero near Brewster's angle for the \textit{p}-polarized incident light. (b) The absorptance as a function of the incident angle and the probe field detuning. The green inset shows the absorption in the absence of the two dielectric layers. (c,d) The reflectance and transmittance as functions of the incident angle and the normalized probe detuning. The thickness of the slabs and atomic medium are $d_{s}=0.5\lambda$, $d_{a}=2\lambda$, respectively. All other parameters remain the same as Fig. \ref{FIG:2}(b).}
\label{FIG:3}
\end{figure}
By inserting this medium in the cavity formed by the dielectric walls, the dissipation of the field can be removed near the resonance, where loss-free enhanced PSHS for the incident p-polarized light can be realized. In Fig. \ref{FIG:3}, we studied the reflectance, transmittance and absorptance using both transfer matrix and finite element simulations. The incident light with \textit{p}-polarization experiences full transmittance when $\theta _{i}\approx \arctan(\eta_i/\eta_0)$ with no dissipation of field inside the cavity. Both the results of theory and FEM are in agreement with each other. The results of simulations are shown by hollow circles in Fig. \ref{FIG:3}(a), and the details of simulations are given in the Appendix. To capture the full response of the intra-cavity medium, we analyzed the absorptance, reflectance, and transmittance against the incident angle $\theta_{i}$ and the normalized probe field detuning $\Delta _{p}$. The green dotted curve is the dissipation (absorption) of the field inside the cavity calculated using Eq. \eqref{eq8}. The two peaks in the absorption of the sole cavity (green curve) aligns that of the whole structure. Apart from the two peaks along the probe field detuning, the absorption in the whole structure completely vanishes at the resonance offering a window for loss-free PSHS enhancement, as shown in Figs. \ref{FIG:3}(b)--(d). 

\begin{figure}
	\centering
	\includegraphics[width=\columnwidth]{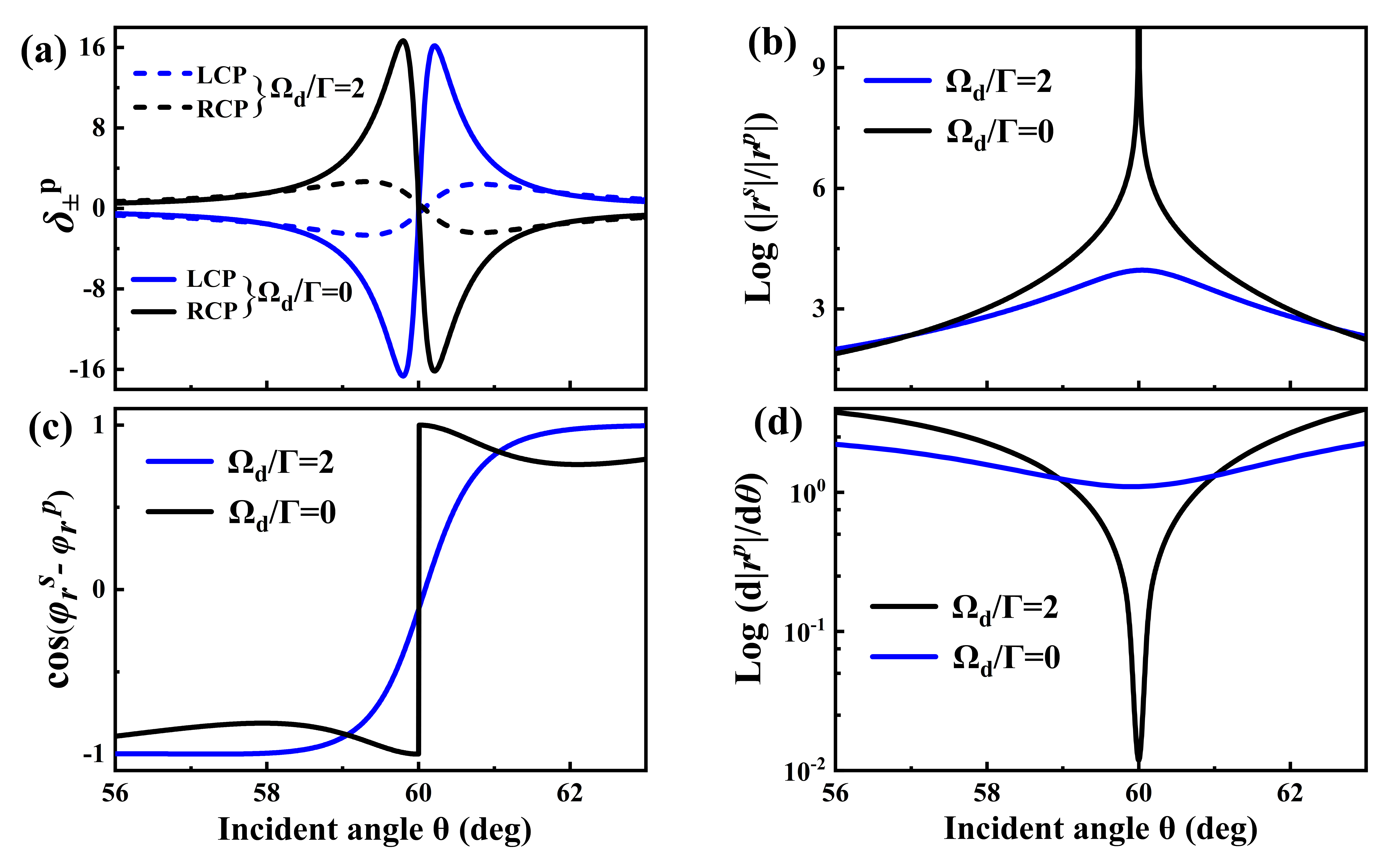}
	\caption{(a) The photonic spin Hall shift (PSHS) for \textit{p}-polarized incident light of the RCP and LCP components in the presence of transparency (solid) and presence of absorption (dotted) at resonance against the incident angle. The PSHS is greatly enhanced by around seven times due to the transparency window, reaching 16 $\lambda$. (b) The log-transformed ratio of the Fresnel's coefficients near Brewster's angle. The onset of transparency at Brewster's angle renders sufficient reduction of the reflectivity for \textit{p}-polarized light. (c) The abrupt phase difference from -1 to 1 at Brewster's angle explains the sign switch of the shift from negative to positive. (d) The log-transformed derivative of the Fresnel's coefficients of \textit{p}-polarized incident light. The magnitude of the derivative is reduced due to induced transparency in the cavity. All the parameters remain the same as Fig. 2(b).}
	\label{FIG:4}
\end{figure}

In Fig. \ref{FIG:4}, we demonstrate the magnitude of PSHS and dig-out the mechanism for the enhancement of PSHS for \textit{p}-polarized incident light. For spin-dependent splitting, we compared the results of EIA and EIT cases. The PSHS is enhanced in the vicinity of Brewster's angle when the cavity is dominated by EIT (solid) rather than EIA (dotted), as illustrated in Fig. \ref{FIG:4}(a). Notably, when $\Omega _{d} $ is absent, PSHS exhibits a sevenfold amplification relative to the $\Omega _{d} $-active condition, and the displacements of LCP and RCP light remain equal in magnitude but opposite in sign with varying incident angles (Fig. \ref{FIG:4}(a)). According to the expression of $\delta _{LCP/RCP}^{p}$, it is obvious that at Brewster's angle, the PSHS diminishes because the Fresnel's coefficient for incident polarization state approaches zero, provided that there is no loss in the cavity. The magnitude of PSHS is exquisitely correlated to the ratio of the Fresnel's coefficients $r^{s}/r^{p} $ and the derivative of the Fresnel's coefficient of \textit{p}-polarized incident light $|\partial r^{p} /\partial \theta _{i} |$. In the presence of the driving field, due to loss in the intracavity medium, the ratio of the Fresnel's coefficients is small and so is the PSHS. However, in the absence of a driving field and the onset of complete transparency, the ratio of Fresnel's coefficient is greatly enhanced while the derivative is reduced near Brewster's angle. This boosts the magnitude of PSHS in the absence of loss. Moreover, from Eq. \eqref{eq6}, one can observe that the PSHS is also proportional to the cosine of the phase difference $\phi=\cos (\varphi _{s} -\varphi _{p} )$ \cite{ling2021revisiting, waseem2024gain}. An abrupt phase jump from -1 to 1 leads to the switch of the sign of the PSHS. Interestingly, this abrupt change arises near Brewster's angle under complete transparency. 
\begin{figure*}
	\centering
	\includegraphics[width=0.8\textwidth]{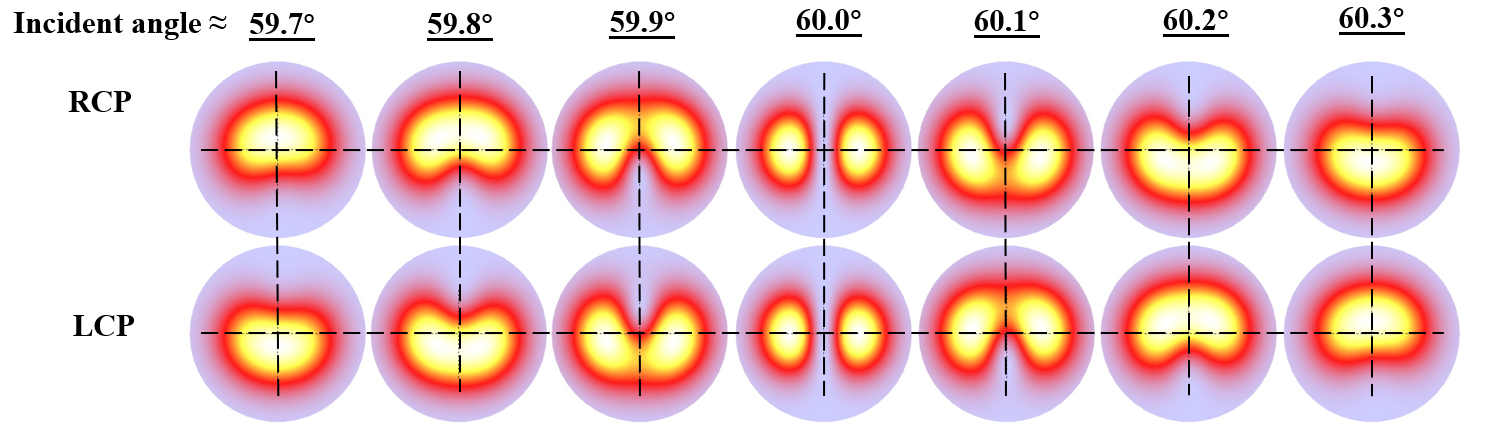}
	\caption{Real space demonstration of photonic spin Hall shift. The beam centroid located at the interface of the first layer and air, evolves from positive to negative when the incident angle exceeds Brewster's angle for the RCP component. The opposite is true for the LCP component. All the parameters remain the same as Fig.\ref{FIG:2}(b).}
	\label{FIG:5}
\end{figure*}

In order to demonstrate the change of the sign at Brewster's angle, we studied the evolution of the reflected beam profile at the interface between air and the first layer in Fig. \ref{FIG:5}. The RCP and LCP components of the reflected beams are shown in the two rows, respectively, for different incident angles near Brewster's angle. The beam waist is taken as 90$\lambda$ and the rest of the parameters are kept the same as in Fig. \ref{FIG:2}(b). It can be seen that in the vicinity of Brewster's angle under induced transparency, the beam profile is symmetric doubled-peak which is essentially a Hermite Gaussian beam with TEM$_{10}$ profile. The PSHS at this angle is nearly zero, as demonstrated by Eq. \eqref{eq6}. On the other hand, the beam centroid changes from positive to negative by crossing Brewster's angle for RCP and conversely, the opposite is true for the LCP component. These results are consistent with the previous results \cite{ling2021revisiting}.

\begin{figure}
	\centering
	\includegraphics[width=\columnwidth]{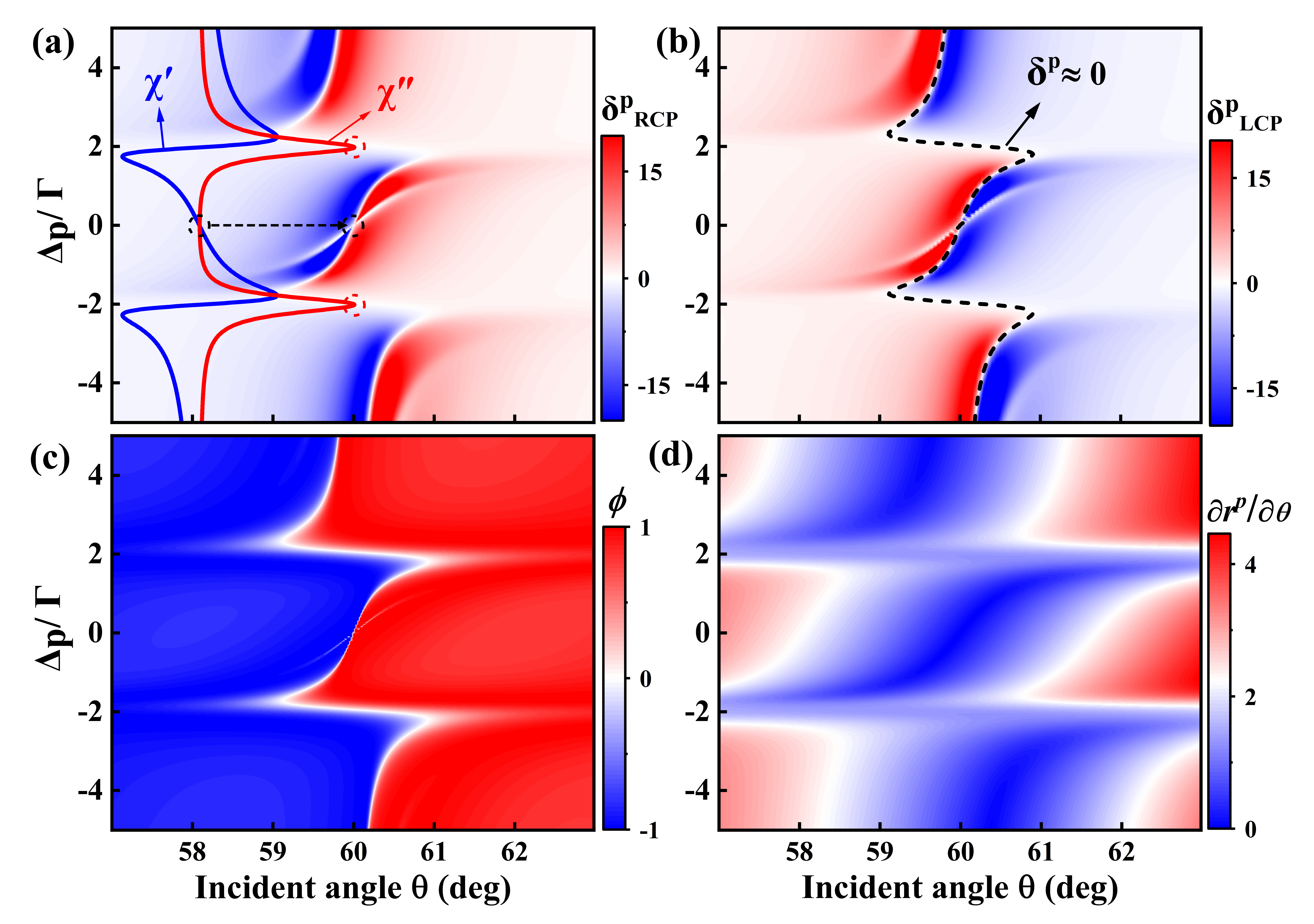}
	\caption{Photonic spin Hall shift against the incident angle and the probe field detuning for (a) RCP and (b) LCP components. The inset shows real and imaginary parts of the complex susceptibility of the medium. The black and red dotted circles shows transparency window and absorption peaks, respectively. (c) Cosine of the phase difference and (d) derivative of the Fresnel's reflection coefficient of the incident \textit{p}-polarization state against the incident angle and the probe field detuning. The black dotted line in (b) shows the regions where the shift approaches zero along which the shift changes its sign due to abrupt phase change. The parameters remain the same as Fig. \ref{FIG:2}(b).}
	\label{FIG:6}
\end{figure}

The PSHS depends on the optical properties of the cavity as well as the refractive index of the dielectric slabs. While the refractive index of the slabs remains fixed, the refractive index of the medium inside the cavity can be precisely controlled by adjusting the strength of the driving fields and the detuning from resonance. Thus, one can directly manipulate the refractive index of the cavity medium to tune the PSHS. To show this, in Fig. \ref{FIG:6}, we investigated the PSHS as a function of both the incident angle and the detuning of the probe field $\Delta _{p}$. For brevity, the real and imaginary parts of the susceptibility of the medium are shown as inset of Fig. \ref{FIG:6}(a). First, the PSHS at exact EIT (inside cavity) and Brewster's angle (black-dotted circle) is nearly zero in exact agreement with Eq. \eqref{eq6} and  Fig. \ref{FIG:5}. By slightly adjusting the detuning of probe field from resonance, the shift in the vicinity of normal dispersion and low absorption is enhanced while it's sign changes due to rapid phase change Fig. \ref{FIG:6}(c). However, as the off-resonant absorption with anomalous dispersion is achieved, the shift is greatly suppressed. This can be explained by the derivative of the Fresnel's reflection coefficient. The derivative of the Fresnel's coefficient of the total structure is low when the absorption is low and vice versa. Secondly, the shift changes it's sign when the one crosses angle of low reflectivity, which can be tuned by the detuning of the probe field as marked by the dotted line in Fig. \ref{FIG:6}(b). Thus, without changing the overall structure to achieve tunability of the PSHS, the proposed model provides a flexible control over the PSHS with low absorption using coherent medium with induced EIT. 

\section{Conclusion} 
In conclusion, we have proposed an approach to enhance photonic spin Hall shift (PSHS) for reflected light. By incorporating a four-level atomic medium which forms a cavity enclosed by two dielectric layers, the optical properties of the medium has been investigated using the Liouville density matrix formulism. Under electromagentic induced transparency (EIT) conditions, the absorption of the medium is completely vanished at resonance to obtain loss-free PSHS near Brewster's angle without introducing compensatory gain mechanisms. Applying the coherent fields to induce transparency in the medium suppresses both the Fresnel's coefficient of the incident light and its derivative for \textit{p}-polarization, thus leading to the enhancement of PSHS. The sign of the enhanced loss-free PSHS changes as a result of the abrupt phase change. Furthermore, the direct manipulation of dispersive properties of the cavity provides a mechanism to tune the PSHS without changing the structure. These findings hold potential for spin-based refractive index sensing applications and cavity quantum electrodynamics devices.

\section{Appendix} 
By calculating the elements of the total transfer matrix, the Fresnel's coefficients of incident linearly polarized light can be obtained by inserting the matrix elements in the following equations:

\begin{widetext}
	\begin{equation}\label{eq9}
		r^{p/s} (\theta _{i} ,d_{j} )=\frac{\cos (\theta )[Q_{22}^{p/s} (\theta ,d_{j} )-Q_{11}^{p/s} (\theta ,d_{j} )]-[\cos ^{2} (\theta )Q_{12}^{p/s} (\theta ,d_{j} )-Q_{21}^{p/s} (\theta ,d_{j} )]}{\cos (\theta )[Q_{22}^{p/s} (\theta ,d_{j} )+Q_{11}^{p/s} (\theta ,d_{j} )]-[\cos ^{2} (\theta )Q_{12}^{p/s} (\theta ,d_{j} )+Q_{21}^{p/s} (\theta ,d_{j} )]} , \tag{A1}  
	\end{equation}
	\vspace{-10pt}
	\begin{equation}\label{eq10}
		t^{p/s} (\theta _{i} ,d_{j} )=\frac{2\cos (\theta )}{\cos (\theta )[Q_{22}^{p/s} (\theta ,d_{j} )+Q_{11}^{p/s} (\theta ,d_{j} )]-[\cos ^{2} (\theta )Q_{12}^{p/s} (\theta ,d_{j} )+Q_{21}^{p/s} (\theta ,d_{j} )]} ,\tag{A2} 
	\end{equation}
\end{widetext}
where $Q_{ij}^{p/s} (\theta _{i} ,d_{j} )$ is the matrix element of the total matrix $Q^{p/s}(\theta _{i} ,d_{j} )$ which can be obtained by the matrix product defined by Eq.\eqref{eq2}. To confirm the results of the transfer matrix, the reflectance and transmittance of the proposed model were numerically simulated using the finite element method in COMSOL Multiphysics. In the transparent regime, both the real and imaginary components of the susceptibility vanish, reducing the system to an equivalent three-layer model, where the middle layer is effectively replaced by an air cavity. To analyze this equivalent model, finite element simulations were performed with air embedded between two slabs, subject to appropriate boundary conditions. The model was constructed as a block composed of three layers, each maintaining the same thickness as in the original structure. Additionally, sufficiently thick air layers were introduced above and below the block to mimic an open-boundary environment. Port boundary conditions were applied: one at the top air layer’s surface for the incident wave and another at the bottom surface to collect the transmitted light, forming a two-port system consistent with the transfer matrix formulation. To mitigate memory constraints associated with large-area simulations, periodic boundary conditions with Floquet periodicity were imposed along the longer parallel sides of the structure. Separate simulations are conducted for \textit{p}- and \textit{s}-polarized light, and the scattering parameters $S_{11}$ and $S_{21}$ are computed to determine the reflectance and transmittance of the proposed model. The results, indicated by hollow circles, are presented in Fig. \ref{FIG:3}(a).

By inputting the Hamiltonian in the rate equation and calculating the various decays, the following matrix elements can be obtained \cite{wan2020controlling}:
\begin{equation}
	\rho _{aa}^{\cdot } =i\Omega _{p} \rho _{ca} -i\Omega _{p} \rho _{ac} +\Gamma _{da} \rho _{dd} +\Gamma _{ca} \rho _{cc} ,  \tag{A3}  
\end{equation}
\begin{equation}
	\rho _{bb}^{\cdot } =i\Omega _{c} \rho _{cd} -i\Omega _{c} \rho _{bc} +i\Omega _{d} (\rho _{db} -\rho _{bd} )+\Gamma _{cb} \rho _{cc} +\Gamma _{db} \rho _{dd}, \tag{A4} 
\end{equation}
\begin{equation}
	\rho _{cc}^{\cdot } =i\Omega _{p} \rho _{ac} -i\Omega _{p} \rho _{ca} +i\Omega _{c} (\rho _{bc} -\rho _{cb} )-(\Gamma _{ca} +\Gamma _{cb} )\rho _{cc},   \tag{A5}
\end{equation}
\begin{equation}
	\rho _{ba}^{\cdot } =i\Omega _{c} \rho _{ca} -i\Omega _{p} \rho _{bc} +i\Omega _{d} \rho _{da} +[(\Delta _{p} -\Delta _{c} )+i\gamma _{ba} ] \rho _{ba}, \tag{A6}   
\end{equation}
\begin{align}
	\rho _{ca}^{\cdot } =&i\Omega _{p} (\rho _{aa} -\rho _{cc} )+i\Omega _{c} \rho _{ba} +i\Omega _{d} \rho _{da} \notag \\&+i[\Delta _{p} +i(\Gamma _{ca} +\Gamma _{cb} )/2]\rho _{ca}, \tag{A7}   
\end{align}
\begin{align}
	\rho _{da}^{\cdot } =&i\Omega _{d} \rho _{ba} -i\Omega _{p} \rho _{dc} +i[(\Delta _{p} -\Delta _{c} +\Delta _{d} )  \notag\\&+i(\Gamma _{da} +\Gamma _{db} )/2] \rho _{da}, \tag{A8}  
\end{align}
\begin{align}
	\rho _{cb}^{\cdot } =&i\Omega _{c} (\rho _{bb} -\rho _{cc} )+i\Omega _{p} \rho _{ba} -i\Omega _{d} \rho _{cd}\notag \\& +i[ \Delta _{c} +i(\Gamma _{ca} +\Gamma _{cb} )/2] \rho _{cb}, \tag{A9}  
\end{align}
\begin{equation}
	\rho _{db}^{\cdot } =i\Omega _{d} (\rho _{bb} -\rho _{dd} )-i\Omega _{c} \rho _{dc} +i[\Delta _{d} +i(\Gamma _{da} +\Gamma _{db} )/2] \rho _{db}, \tag{A10}   
\end{equation}
\begin{align}
	\rho _{dc}^{\cdot } =&i\Omega _{d} \rho _{bc} -i\Omega _{p} \rho _{da} -i\Omega _{c} \rho _{db} +i[(\Delta _{c} -\Delta _{d} )\notag \\&+i(\Gamma _{ca} +\Gamma _{cb} +\Gamma _{da} +\Gamma _{db} )/2] \rho _{dc}. \tag{A11}   
\end{align}
\vspace{\baselineskip}
\section*{Acknowledgments}
This work was supported by National Natural Science Foundation of China (12174281, 12274314, 12311530763, 12374273, 12421005); Hunan Provincial Major Sci-Tech Program (2023ZJ1010); Natural Science Foundation of Jiangsu Province (BK20221240); Suzhou Basic Research Project (SJC2023003).
\renewcommand{\bibname}{References}

\bibliography{ref.bib}
\end{document}